\documentclass{pasj01}
\Received{$\langle$reception date$\rangle$}
\Accepted{$\langle$acception date$\rangle$}
\Published{$\langle$publication date$\rangle$}
\usepackage{longtable}
\usepackage{threeparttable}
\usepackage[figuresright]{rotating}
\usepackage[switch,mathlines]{lineno}
\begin{document}

\title{ MK-like spectral classification for hot subdwarf stars with LAMOST spectra}

\author{Xuan \textsc{Zou},\altaffilmark{1,2}}
\author{Zhenxin \textsc{Lei},\altaffilmark{1,2*}}
\altaffiltext{1}{Key Laboratory of Stars and Interstellar Medium, Xiangtan University, Xiangtan 411105, People’s Republic of China}
\altaffiltext{2}{Physics Department, Xiangtan University, Xiangtan, 411105, People’s Republic of China}
\email{leizhenxin2060@163.com}

 
\maketitle
\begin{abstract}
An MK-like spectral classification has been  conducted for 1224 hot subdwarf stars with LAMOST DR9 low-resolution spectra. The whole sample was divided into four categories according to the spectral line characteristics: He-normal, He-weak, He-strong C and He-strong. Each selected spectrum was assigned a spectral class, a luminosity class and an helium class by comparing the line depth and width with standard spectra selected in LAMOST. Relationships between atmospheric parameters and spectral classification were also presented. 
\end{abstract}

\section{Introduction}
Hot subdwarf stars locate in the region between main-sequence (MS) and White Dwarf (WD) cooling curve in the Hertzsprung-Russell (H-R) diagram \citep{2009ARA&A..47..211H, 2016PASP..128h2001H}. Many of them are core helium (He) burning stars with very thin hydrogen (H) envelopes (e.g., less than 0.02 M$_{\odot}$). Some of them even evolve off He burning stage and are on the way toward WD cooling curve. Hot subdwarf stars in globular clusters (GCs) locate on the extreme horizontal branch (EHB) and known as EHB stars. 
   
Hot subdwarfs play significant roles in many research aspects of astrophysics, especially in binary evolution. \citet{2002MNRAS.336..449H} and \citet{2003MNRAS.341..669H} conducted binary population synthesis to study formation of sdB type hot subdwarf stars. They found that stable Roche lobe overflow (RLOF), common envelope (CE) ejection and the merger of two He-WDs in binary evolution can produce sdB stars, and most of the properties of these stars predicted by models, such as  period and mass distribution, positions in the $T_{\rm eff}$ - $\log{g}$ plane are consistent with observations. \citet{2023arXiv231117304G} studied the efficiency of CE ejection in binary evolution by using short period sdB+WD binaries. 

Hot subdwarfs could be main sources of ultraviolet (UV) upturn phenomena found in elliptical galaxies \citep{1999ARA&A..37..603O, 2007MNRAS.380.1098H}. Hot subdwarfs with massive WD companions are considered as progenitors of type Ia supernovae (SNe Ia, \cite{2017A&A...600A..50G, 2018RAA....18...49W, 2023RAA....23h2001L}). In addition,  binaries consisting of hot subdwarfs and WDs are possible gravitational wave (GW) sources  \citep{2006astro.ph.12535O,2013A&A...557A.122G,2022ApJ...933..137G, 2018MNRAS.480..302K, 2024ApJ...963..100K}. Pulsating hot subdwarfs can be used to study the inner structure of these stars through astroseismology method \citep{2011A&A...530A...3C}.

Recently, a large number of new hot subdwarfs have been discovered thanks to the new data releases from large photometric and spectroscopic surveys, such as Sloan Digital Sky Survey  Data Release 14 \citep{2018ApJS..235...42A, 2019MNRAS.486.2169K}, the third data release of Gaia mission \citep{2023A&A...674A...1G},  The Large Sky Area Multi-Object Fiber Spectroscopic Telescope (LAMOST) spectral survey \citep{2018ApJ...868...70L, 2019ApJ...881..135L,2020ApJ...889..117L, 2023ApJ...942..109L,2021ApJS..256...28L}, and The Transiting Exoplanet Survey Satellite (TESS, \cite{2021A&A...651A.121U,2022A&A...663A..45K,2022A&A...666A.182S,2022MNRAS.516.1509K, 2023A&A...673A..90S,2023MNRAS.519.2486S}). \citet{2017A&A...600A..50G} compiled a catalog of known hot subdwarf from  literature and published tables, in which 5613 objects are included. With newly discovered  hot subdwarfs increasing quickly, the number of known hot subdwarfs was updated to 5874 by \citet{2020A&A...635A.193G} and 6616 by \citet{2022A&A...662A..40C}, respectively. 

According to the characteristic of spectral lines, hot subdwarf stars can be classified into sdB, sdOB, sdO, and their He-rich counterparts He-sdB, He-sdOB, He-sdO \citep{1990A&AS...86...53M, 2017A&A...600A..50G, 2018ApJ...868...70L}. \citet{2023ApJ...953..122L} presented mass distributions for 664 single-lined hot subdwarf stars, and analyzed the possible formation channels for these stars with different spectral classifications. However, this kind of spectral classification scheme cannot reflect directly the differences of temperature, luminosity and surface He abundance among stars with similar spectral type. Therefore, a more detailed MK-like classification system needs to be developed for these stars \citep{1996ASPC...96..461D}. \citet{1997A&AS..125..501J} introduced a classification scheme for He-rich hot subdwarfs, which can be extended to all hot subdwarfs with spectra dominated by H, HeI, or HeII lines. Furthermore, \citet{2013A&A...551A..31D} proposed a new three-dimensional MK-like classification method for hot subdwarf stars, which gives proxies for effective temperature (spectral class), surface gravity (luminosity class), and surface He abundance (He class). By employing this classification scheme,  \citet{2019PASJ...71...41L} classified 56 hot subdwarfs stars identified in LAMOST DR1, and \citet{2021MNRAS.501..623J} classified 107 He-rich  hot subdwarfs observed by Southern African Large Telescope (SALT). 

 Although  more than 6000 hot subdwarf stars have been discovered to date \citep{2022A&A...662A..40C}, few of them have MK-like spectral classifications. In this study, employing the MK-like spectral classification scheme  designed by \citet{2013A&A...551A..31D}, we classified the hot subdwarfs recorded in \citet{2022A&A...662A..40C} with low-resolution spectra of LAMOST DR9. In Section 2, we introduced the sample selection and how to selected standard stars  in LAMOST. The results and discussion were given in Section 3, and a brief conclusion is give in Section 4.

\section{Sample selection and standard stars}

\subsection{Sample selection}

LAMOST is a 4m specially designed Schmidt survey telescope \citep{2012RAA....12.1197C, 2012RAA....12..723Z} at Xinglong Station of the National Astronomical Observatories of  Chinese Academy of Sciences (CAS). It can simultaneously collect spectra of 4000 objects in a field of view 20 deg$^{2}$. LAMOST begun it's pilot survey in October 2011 and ended in June 2012. Until July 2021, LAMOST finished regular survey observation for the first nine year. The data collected both in pilot and regular survey for the first nine year were published as the ninth data release (DR9). 
LAMOST DR9 dataset contains 11,211,028 spectra, in which 10,893,354 stellar spectra, 241,454 galaxy spectra, and 76,220 quasar spectra are included. These spectra cover a wavelength range of 3700 - 9000 \AA \ with a resolution of $\lambda/\Delta\lambda=$ 1800 at 5500 \AA.

\citet{2022A&A...662A..40C} compiled a new catalog for known hot subdwarf stars, in which 6616 objects are included. Among these hot subdwarf stars, 3087 stars have atmospheric parameters, while 2791 stars have radial velocity values. Furthermore, they also compiled a catalog for hot subluminous stars based on the Early Data Release 3 of Gaia mission (Gaia EDR3, \cite{2021A&A...649A...1G}), in which more than 60,000 hot subdwarf candidates were collected. The two catalogs provide great convenience to study these blue stars in our galaxy. 

We cross-matched LAMOST DR9 low-resolution dataset with the catalog of known hot subdwarf stars compiled by \citet{2022A&A...662A..40C}, and obtained 1922 common objects. Then, we selected objects for following analysis with their spectral signal-to-noise ratios in the $u$ band (SNRU) greater than 10. After removing duplicated sources, we finally selected 1252 hot subdwarfs for MK-like spectral classification. 

\subsection{MK-like classification system}
\begin{figure}
	\includegraphics[width=80mm]{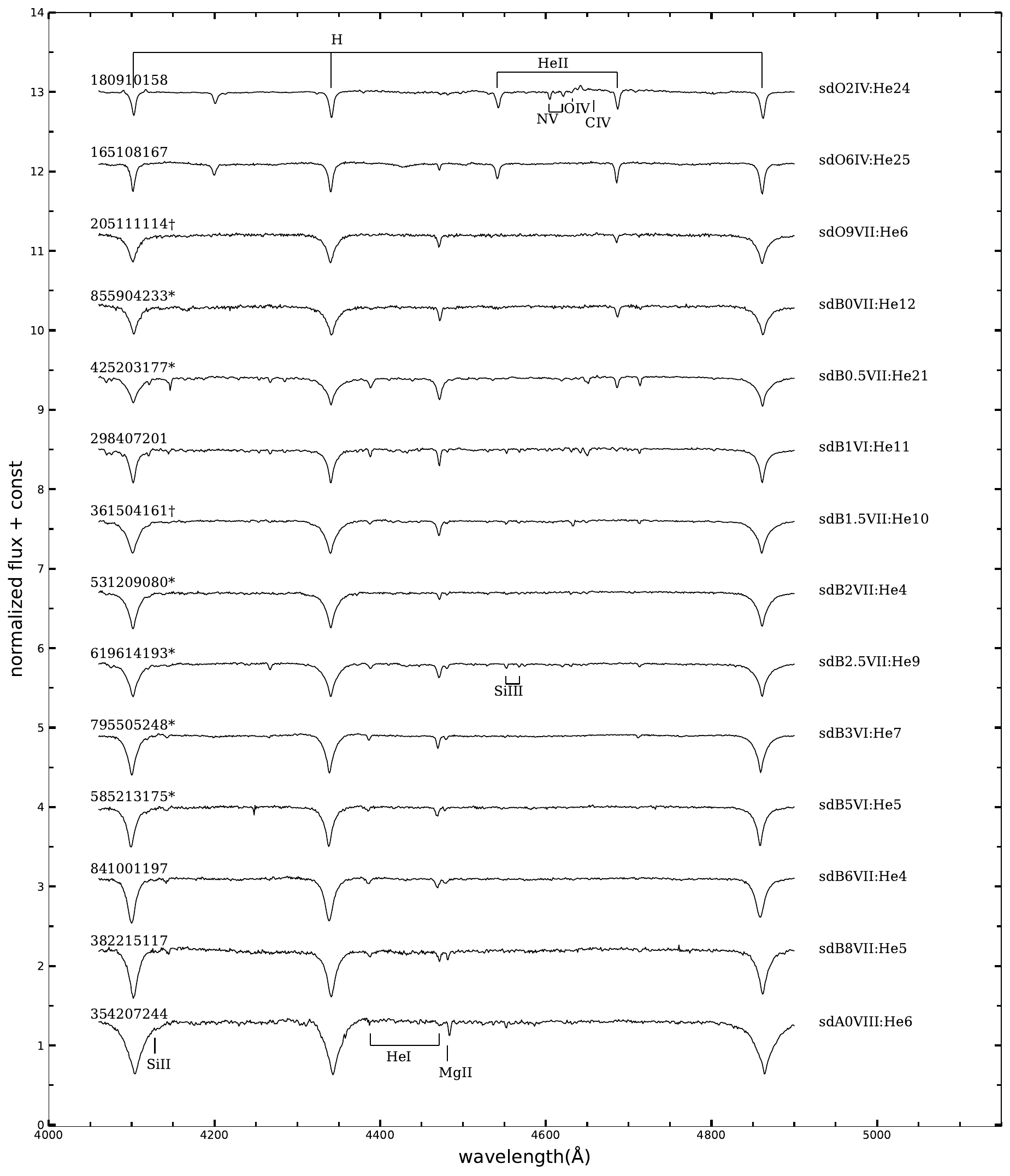}
    \caption{Selected He-normal standard spectra from LAMOST. Primary spectral lines labeled are H Balmer lines $\lambda\lambda$4101,4340,4861, HeI lines $\lambda\lambda$4387,4471, HeII lines $\lambda\lambda$4541,4686 and metallic lines SiII $\lambda\lambda$4128, MgII $\lambda\lambda$4481, SiIII $\lambda\lambda$4552-68, OIV $\lambda\lambda$4632, NV $\lambda\lambda$4604–20, CIV $\lambda\lambda$4658 blend. LAMOST obsid and spectral classification for each spectrum are  shown at left and right of the panel, respectively.  }
    \label{fig1}
\end{figure}

The spectral classification scheme in \citet{2013A&A...551A..31D} (hereafter D13) consists of a spectral class, a luminosity class and an Helium  class. To be assigned a detailed spectral class, hot subdwarf stars should be divided into four categories: (1) He-normal category: spectra present dominated H lines, but metallic lines, HeI and HeII lines are visible. (2) He-weak category: spectra also present dominated H lines, either HeI or HeII lines can exist (but not both). Metallic lines are very weak or absent. (3) He-strong C category: spectra present strong He lines, and carbon lines are also present. (4) He-strong category: spectra also present strong He lines, but carbon lines are very weak or absent. 

In each category, hot subdwarf stars are arranged into sequences where the depth of characteristic lines vary in a gradual manner. After that, standard spectra will be selected among the sequences (see Section 2.3 for detailed description).  Finally, spectral class for all hot subdwarf stars can be obtained by interpolating the line depth of each spectrum with selected standard stars. 

On the other hand, luminosity class is determined by comparing the line width (e.g., H$_{\gamma}$ or He lines) of each spectrum  with average width of the same line for all stars. If line width of a spectrum is close to the average line width of all sample, then a standard luminosity class VII for hot subdwarf stars will be assigned. If line width is noticeably broader than the average value, then a class VIII is assigned. While if the line width is obviously narrower than the average value, a class VI is assigned. For spectra with line widths narrower than the spectra with class VI, luminosity classes I – V will be assigned by comparing with O/B standard stars in MK classification system \citep{2009ssc..book.....G}. The line width for each spectrum is obtained by fitting the profile of specific lines (see Section 2.4 for detailed description).

As described in D13, He class is denoted by an integer number between 0-40, which indicates the strengths of He lines relative to H lines. The He  class 0 denotes that no He lines are shown in the spectrum, while He class 40 denotes that no H lines are shown. Specifically, for He class 0-20,  the integer number is roughly equal to the following function of line depth:
\begin{equation}
    20*(d_{\rm 4471}+d_{\rm 4541})/(d_{\rm \gamma}-0.83d_{\rm 4541}), 
	\label{eq:1}
\end{equation}
Where $d_{\rm 4471}$, $d_{\rm 4541}$ and $d_{\rm \gamma}$ are line depth of HeI $\lambda\lambda$ 4471, HeII $\lambda\lambda$ 4541 and H$_{\gamma}$, respectively.  
While for He  class 20 - 40, the number is roughly calculated by the following function of line depth: 
\begin{equation}
40-20*(d_{\rm \gamma}-0.83d_{\rm 4541})/(d_{\rm 4471}+d_{\rm 4541})
	\label{eq:2}
\end{equation}

\subsection{Standard stars}

\begin{figure}
	\includegraphics[width=80mm]{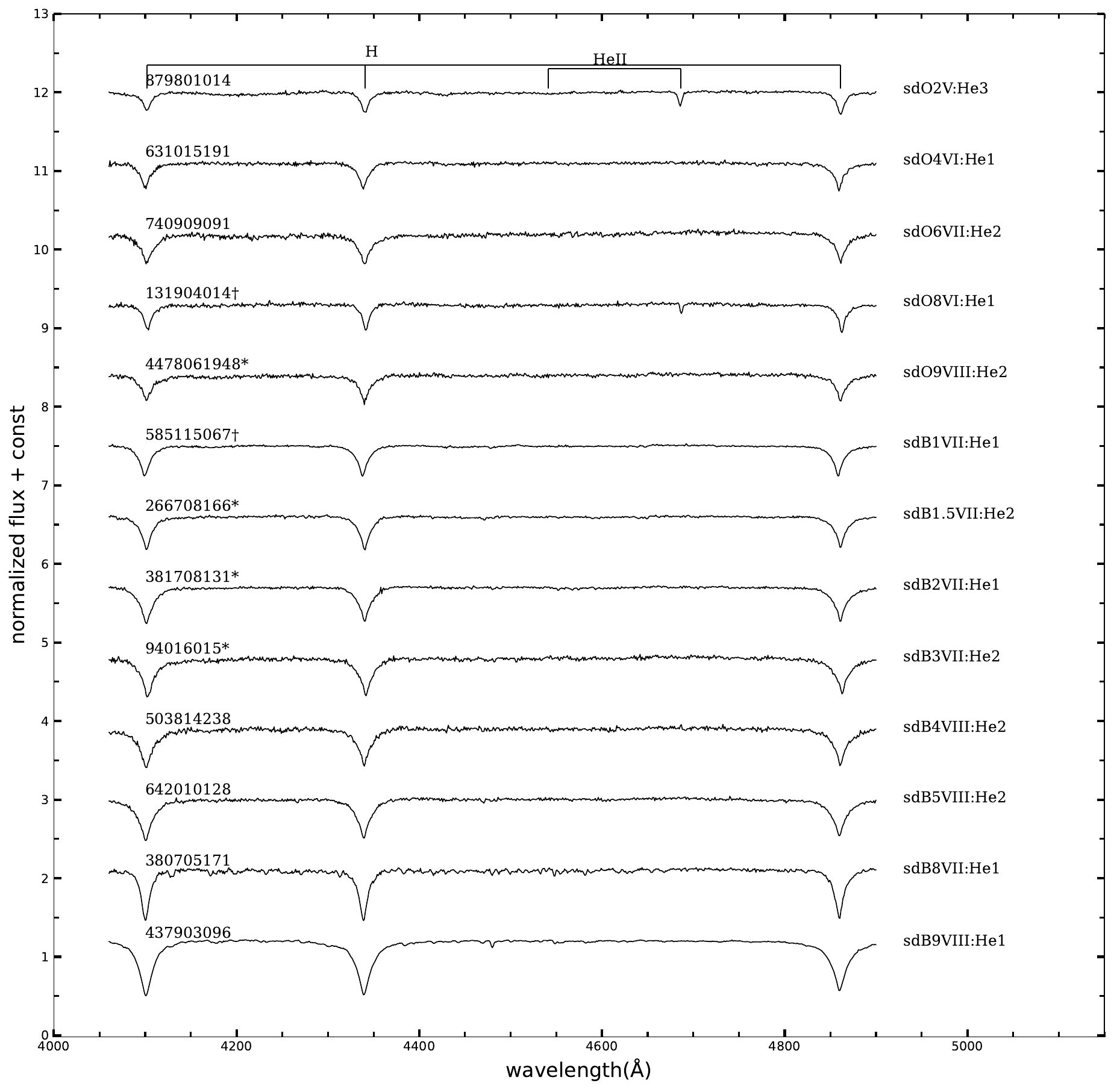}
    \caption{Selected He-weak standard spectra from LAMOST. Primary spectral lines labeled are H + HeII lines $\lambda\lambda$ 4101,  4340, 4861, 4541, 4686. LAMOST obsid and spectral classification for each spectrum are  shown at left and right of the panel, respectively.}
    \label{fig2}
\end{figure}

The spectra analyzed in D13 are mostly from Palomar-Green (PG) survey \citep{1986ApJS...61..305G}, which have a resolution of $\lambda/\Delta\lambda \approx$ 2000 within wavelength range of 4050 - 4900 \AA . However, the resolution of LAMOST spectra (e.g., $\lambda/\Delta\lambda \approx$ 1800 at 5500 \AA) is a little lower than the spectra in PG survey. Since spectral resolution and signal-to-noise ratio (SNR) would influence spectral classification, thus accurate spectral classification should be done if the standard stars are observed with the same instrument in order to eliminate systematic errors. Therefore, we need to select our own standard stars when trying to classify hot subdwarf stars with  LAMOST spectra. 

 To be consistent with the standards adopted in D13, we cross-matched the standard stars of D13 with LAMOST spectra, and found 17 common objects,  among which 8 of them have good quality (e.g., SNRU$\geq$10.0) of LAMOST spectra. These stars are selected as the main standards in this study. Additionally, some stars classified in D13 and \citet{2021MNRAS.501..623J}, which have good quality of LAMOST spectra, are also selected as standard stars. We also selected visually some stars in LAMOST as standards by comparing them with the spectra classified in D13 (See Table 1 for details). 
   
\begin{figure}
	\includegraphics[width=80mm]{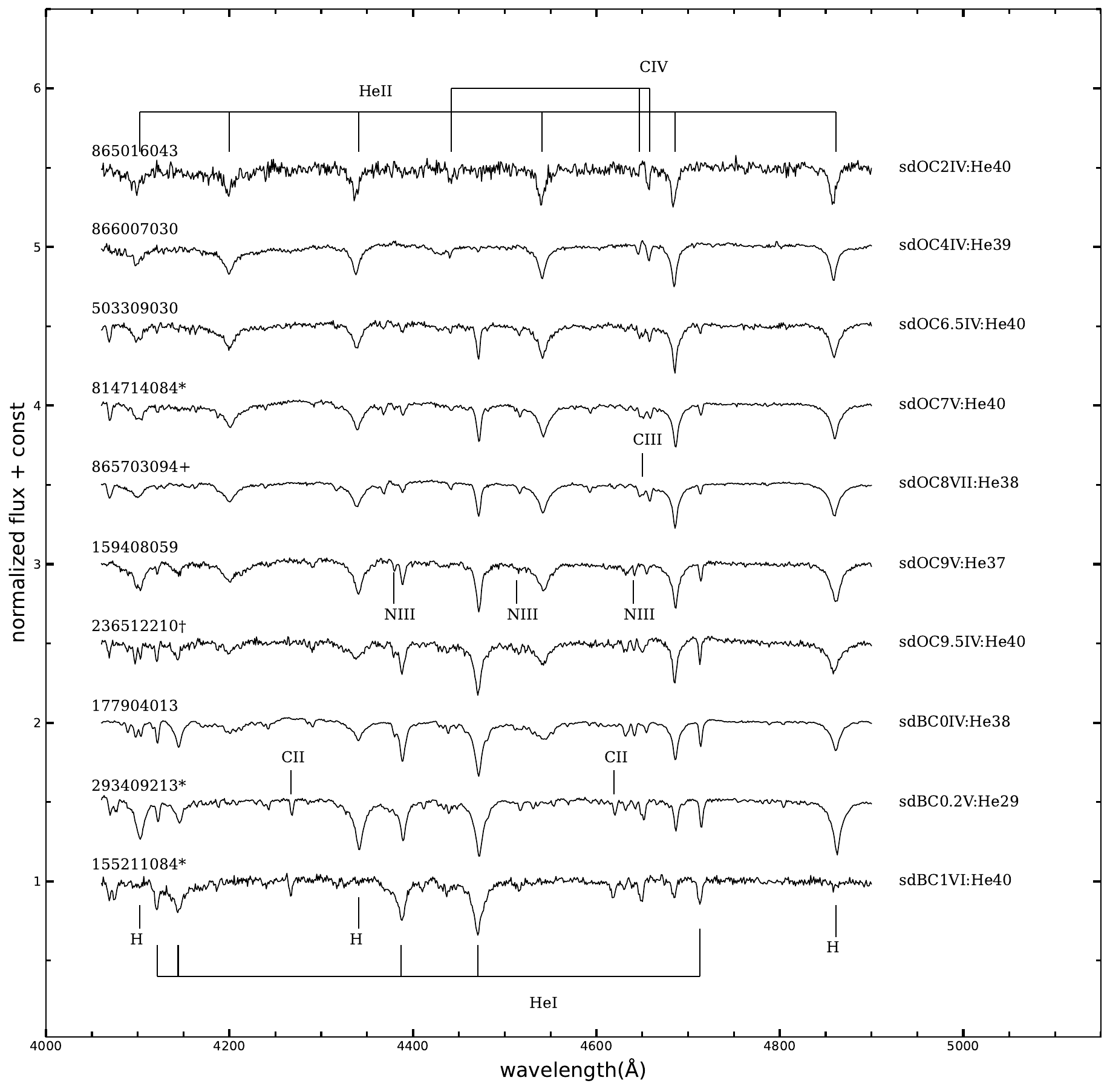}
    \caption{Selected He-strong C standard spectra from LAMOST. Primary spectral lines labeled are HeII $\lambda\lambda$ 4100, 4200, 4338, 4541, 4686, 4859,  CIV $\lambda\lambda$ 4442, 4647, 4658, CIII $\lambda\lambda$ 4650 blend, NIII $\lambda\lambda$ 4379, 4511–15, 4640 blend ,  HeI $\lambda\lambda$ 4121, 4144, 4387, 4471,4713,  CII $\lambda\lambda$ 4267, 4619, H $\lambda\lambda$ 4101, 4340, 4861. LAMOST obsid and spectral classification for each spectrum are shown at left and right of the panel, respectively.}
    \label{fig3}
\end{figure}

\begin{figure}
 \begin{center}
\includegraphics[width=80mm]{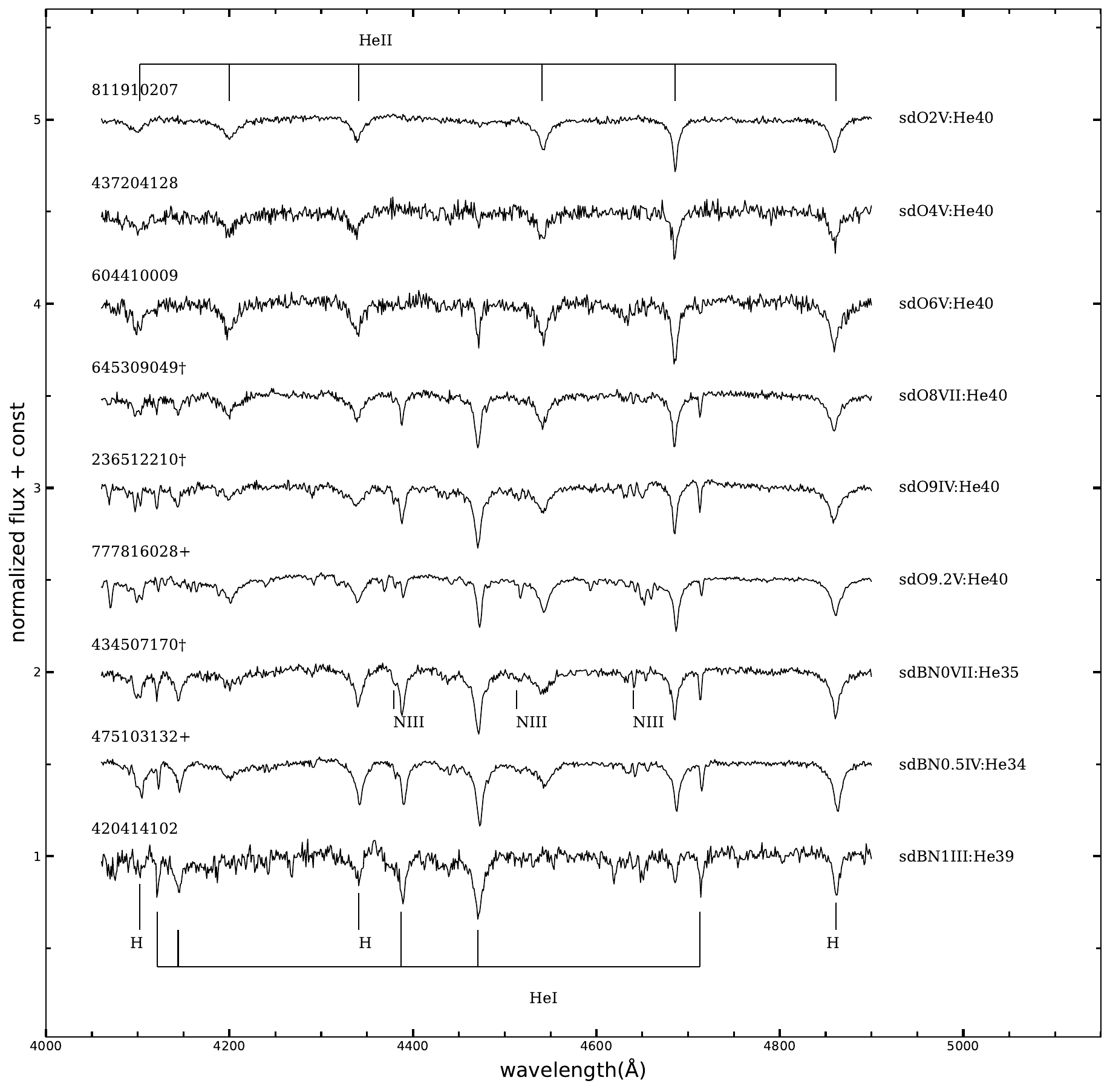} 
\end{center}
\caption{Selected He-strong standard spectra from LAMOST. Primary spectral lines labeled are HeII $\lambda\lambda$ 4100, 4200, 4338,4541, 4686, 4859, HeI $\lambda\lambda$ 4121, 4144, 4387, 4471,4713, NIII $\lambda\lambda$ 4379, 4511–15, 4640 blend, H $\lambda\lambda$ 4101, 4340, 4861. LAMOST obsid and spectral classification for each spectrum are shown at left and right of the panel, respectively. }
\label{fig4}
\end{figure}

The selected hot subdwarf stars in LAMOST were divided into four categories as described in Section 2.2. Then, all the spectra were arranged in sequences for each category, where line depths and widths vary in a smooth manner from spectrum to spectrum. The standard stars selected in this study were listed in Table \ref{tab:1}.  Stars labeled with '$\dag$' denote the standard stars used in D13, and stars labeled with '*' are hot subdwarf stars classified in D13 with good quality of LAMOST spectra, while stars labeled by '+' are He-rich hot subdwarf stars classified in \citet{2021MNRAS.501..623J}. The spectra of all standard stars  for each category are presented in Fig \ref{fig1} – \ref{fig4} respectively.  

 He-normal standard spectra were shown in Figure \ref{fig1} \footnote{LAMOST 180910158 and 165108167 have He classes of 24 and 25 respectively, which indicate intermediate He abundance for the two stars. However, to keep consistence with the spectral classification scheme of D13, we still classified them as He-normal stars. See the definition of He-normal stars in Section 3 and Fig 1 of D13.}. The spectral class of spectra for this category were assigned by comparing them with selected standard stars, which is primarily according to the ratio of HeII $\lambda\lambda$4542/HeI $\lambda\lambda$4471. However, He II $\lambda\lambda$4541 could be weak or absent for some hot subdwarf stars, especially for spectral class earlier than B1. In such cases, the ratio of HeI $\lambda\lambda$4471/HeII $\lambda\lambda$4686 can be used as an alternative criterion. In general, stars with more earlier spectral class often exhibit stronger ionized helium lines, such as HeII $\lambda\lambda$4541 and HeI $\lambda\lambda$4471. On the other hand, as for stars with later spectral class, the relative strength of the helium lines decrease.  

 He-weak standard spectra were shown in Figure \ref{fig2}. Spectra in this category present dominate H lines, together with either HeI or HeII  lines (but not both). These spectra were arranged according to the depth of Balmer lines. If Balmer line depth of a spectrum is consistent with the spectrum shown in Figure \ref{fig1}, they will be assigned the same spectral class. 

He-strong C standard spectra were shown in Figure \ref{fig3}. Unlike the previous two categories, all spectra in this category were arranged according to the depths of Pickering lines. The spectral class of observed spectra can be assigned based on relative strengths of Pickering lines by comparing them with standard spectra in this category. From sdOC2 to sdBC1, line depth ratio of He I$\lambda\lambda$ 4471/He II$\lambda\lambda$4541 was used as the primary criterion. In addition, from sdOC6.5 to sdBC1, the ratio of He I $\lambda\lambda$4713/He II $\lambda\lambda$4686  and the ratio of He I $\lambda\lambda$4471/He II $\lambda\lambda$4686 were used as the secondary criterion for the classification in this category. Since carbon lines are strong in the spectra of this category, a label of "C" is added in to the spectral class to distinguish them with the spectra from other categories.

 He-strong standard spectra were shown in Figure \ref{fig4}. Spectra in this category also present dominate He lines as in He-strong C category, but with no obvious C lines. A label of  "N" will be added into spectral class in this category if NIII lines are visible in the spectra. It can help differentiate them from the stars in He-strong C category.

\begin{table}
 \caption{Standard stars selected in LAMOST.}
 \label{tab:1}
 \fontsize{8pt}{10pt}\selectfont
 \begin{tabular*}{82mm}{l@{\hspace*{10pt}}l@{\hspace*{8pt}}l}
 \\
 \hline
  LAMOST\_obsid & Name & Classification \\ 
  \hline
  \multicolumn{3}{c}{He-normal} \\
\hline
180910158& - &sdO2IV:He24\\
165108167& - &sdO6IV:He25\\
205111114$^\dag$& PG1047+003 &sdO9VII:He6\\
855904233$^*$&  PG0921+161 &sdB0VII:He12\\
425203177$^*$& PG0909+276 &sdB0.5VII:He21\\
298407201& - &sdB1VI:He11\\
361504161$^\dag$& PG2337+070 &sdB1.5VII:He10\\
531209080$^*$& PG0907+123 &sdB2VII:He4\\
619614193$^*$& PG0004+133 &sdB2.5VII:He9\\
795505248$^*$& PG1451+492 &sdB3VI:He7\\
585213175$^*$& PG1705+537 &sdB5VI:He5\\
841001197& - &sdB6VII:He4\\
382215117& - &sdB8VII:He5\\
354207244& - &sdA0VIII:He6\\
\hline
\multicolumn{3}{c}{He-weak}\\
\hline
879801014& - &sdO2V:He3\\
631015191& - &sdO4VI:He1\\
740909091& - &sdO6VII:He2\\
131904014$^\dag$& PG1017+431 &sdO8VIHe1\\
447806194$^*$& PG1538+401 &sdO9VIIIHe2\\
585115067$^\dag$& PG1648+536 &sdB1VIIHe1\\
266708166$^*$& HS0016+0044 &sdB1.5VIIHe2\\
381708131$^*$& PG2205+023 &sdB2VIIHe1\\
94016015$^*$& PG0856+121 &sdB3VIIHe2\\
503814238& - &sdB4VIII:He2\\
642010128& - &sdB5VIII:He2\\
380705171& - &sdB8VII:He1\\
437903096& - &sdB9VIII:He1\\
\hline
\multicolumn{3}{c}{He-strong C}\\
\hline
865016043& - &sdOC2IV:He40\\
866007030& - &sdOC4IV:He39\\
503309030& - &sdOC6.5IV:He40\\
814714084$^*$& PG0838+133 &sdOC7V:He40\\
865703094$^+$& [CW83] 0904-02 &sdOC8VII:He38\\
159408059& - &sdOC9V:He37\\
236512210$^\dag$& PG1624+085 &sdOC9.5IV:He40\\
177904013& - &sdBC0IV:He38\\
293409213$^*$& PG0240+046 &sdBC0.2V:He29\\
155211084$^*$& PG1544+488 &sdBC1VI:He40\\
\hline
\multicolumn{3}{c}{He-strong}\\
\hline
811910207& - &sdO2V:He40\\
437204128& - &sdO4V:He40\\
604410009& - &sdO6V:He40\\
645309049$^\dag$& PG1325+054 &sdO8VII:He40\\
236512210$^\dag$& PG1624+085 &sdO9IV:He40\\
777816028$^+$& GALEX J042034.8+012041 &sdO9.2V:He40\\
434507170$^\dag$& PG0921+311 &sdBN0VII:He35\\
475103132$^+$& GALEX J075807.5-043203 &sdBN0.5IV:He34\\
420414102& - &sdBN1III:He39\\
  \hline
 \end{tabular*}
 \begin{tablenotes}
 \footnotesize
\item[\dag] Stars labeled by '$\dag$' are common objects in D13 standard stars.
\item[*] Stars labeled by '*' are common objects classified in D13.
\item[+] Stars labeled by '+' are common objects classified in \citet{2021MNRAS.501..623J}.

 \end{tablenotes}
\end{table}

\subsection{Spectral line profile fitting}

\begin{figure}
	\includegraphics[width=70mm]{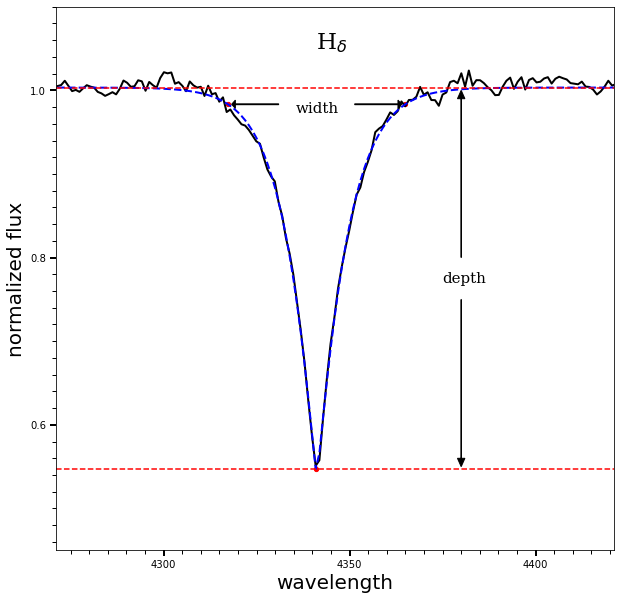}
    \caption{An example of line profile fitting for $\rm{H}_\delta$. Black solid curve is line profile of observed spectrum, while blue dashed curve is the fitting line profile. Line depth and width are indicated by black arrows. }
    \label{fig5}
\end{figure}

To measure width and depth of specific lines for each spectrum, a scale width versus shape method was employed in this study \citep{2002MNRAS.337...87C,2008ApJ...684.1143X, 2019PASJ...71...41L}. It provides a more objective and standardized approach for spectral classification. This method is based on a Sersic profile fit  \citep{1968adga.book.....S} to Balmer or Pickering lines in the following form:
   \begin{equation}
    y=1.0-a\times\ \rm{exp}[-(\frac{|\lambda-\lambda_{\rm 0}|}{b} )^c]
	\label{eq:3}
   \end{equation}
where y is normalized flux, $\lambda$ is  wavelength, and $\lambda_{0}$ is nominal central wavelength of Balmer or Pickering lines. The coefficients a, b, and c are free parameters to be fixed. To account for imperfect normalization of the selected spectra, five free parameters were used in this study to fit the profile of each normalized spectrum, e.g., a, b, c, $\lambda_{0}$, and n,
   \begin{equation}
    y=n-a\times \ \rm{exp}[-(\frac{|\lambda-\lambda_{\rm 0}|}{b} )^c].
	\label{eq:10}
   \end{equation}
   
Fig \ref{fig5} gives an example of line profile fitting for $\rm{H}_{\delta}$. As shown in the figure, line width is defined at the position which is 0.02 below the local continuum, while line depth is defined as the normalized flux relative to the continuum at line core. Note that values of width and depth for a few stars can not be obtained or not reliable due to bad  quality of spectra , therefore their spectral classifications will not be reported in this study. With the line depths measured through this method, He class can be calculated directly from Equation \ref{eq:1} and Equation \ref{eq:2} for each selected sample. Furthermore, spectral classes for all the selected stars can be obtained by interpolating visually the line depths with standard spectra, and luminosity classes can be obtained by using the line width (see Section 2.2). 

\section{Results and Discussion} 
\subsection{Results of spectral classification} 

Employing the method described in Section 2, 1224 selected hot subdwarf stars have been classified successfully, but 28 stars were not classified due to bad spectra quality. Among 1224 classified stars, 702 of them belong to He-normal category, 396 stars belong to He-weak category, 81 stars belong to He-strong C category, and 45 stars belong to He-strong category. 

Table \ref{tab:2}\footnote{This table lists only a portion of the 1224 stars. The whole catalog will be uploaded to CDS after publication of this paper).} presents main properties of 1224 hot subdwarf stars together with the MK-like spectral classification obtained in this study. Columns 1 - 4 give obs\_id, RA, DEC, SNRU in LAMOST, columns 5-6 give Gaia G magnitudes and their uncertainties. Columns 7 - 12 give $T_{\rm eff}$, $\log{g}$, surface He abundance (i.e., $\log(n{\rm He}/n{\rm H})$) and their uncertainties reported in \citet{2022A&A...662A..40C} (labeled by Culpan22). Columns 13 - 15 give spectral class, He class and luminosity class obtained in this study and the columns 16 - 18 give their numerical form. The stars labeled by '$\ast$' and '+' are common objects in D13 and \citet{2021MNRAS.501..623J}, respectively (see Section 3.2 for detailed discussion). 

\onecolumn
\begin{sidewaystable}[ht]
 \fontsize{5.5pt}{9pt}\selectfont
\caption{Main parameters for the 1224 hot subdwarfs with MK-like spectral classifications in this study.} 
\centering 
 \resizebox{\textwidth}{!}{
\begin{tabular}{lccccccccccccccccc} 
\hline\hline 
obs\_id & RA(J2000) & DEC(J2000) & SNRU & GGaia & e\_GGaia & $T_\mathrm{eff}$ & e\_$T_\mathrm{eff}$ & $\mathrm{log}\ g$ & e\_$\mathrm{log}\ g$ & $\log(n{\rm He}/n{\rm H})$ & e\_$\log(n{\rm He}/n{\rm H})$  & spectral class & He-class & luminosity class & spectral\_num & He\_num & luminosity\_num\\
LAMOST & LAMOST & LAMOST & LAMOST & Gaia & Gaia & Culpan22 & Culpan22 & Culpan22 & Culpan22 & Culpan22 & Culpan22 & this study & this study & this study & this study & this study & this study\\
(1)&(2)&(3)&(4)&(5)&(6)&(7)&(8)&(9)&(10)&(11)&(12)&(13)&(14)&(15)&(16)&(17)&(18)\\[0.2ex]
\hline 
472903166&0.065689353&17.64803258&36.71&16.5199&0.0029&38462&184&6.02 &0.03 &-1.92 &0.05 &sdO9.7&He9&VII&0.97 &9&7\\
385816105&0.076711055&22.30076813&38.71&14.2196&0.0028&27629&500&5.55 &0.05 &-2.54 &0.05 &sdB0&He5&VII&1.00 &5&7\\
66605009*&0.278022646&11.01003763&21.55&13.5899&0.0028&27939&216&5.49 &0.03 &-2.73 &0.03 &sdB4&He5&VIII&1.40 &5&8\\
385805170&0.535270916&19.98698107&15.23&15.5813&0.0029&31434&500&5.56 &0.05 &-3.58 &0.05 &sdO5&He3&VII&0.50 &3&7\\
492112073*&0.981704155&27.81032196&85.51&13.3089&0.0028&25400&500&5.30 &0.10 &-2.87 &0.10 &sdB4&He4&VIII&1.40 &4&8\\
385811215&1.02538595&23.03013&24.71&14.8681&0.0028&42494&576&5.44 &0.04 &-4.04 &0.26 &sdO2&He1&V&0.20 &1&5\\
271008159&1.165420485&41.54278088&46.33&15.1129&0.0028&-&-&-&-&-&-&sdO2&He2&V&0.20 &2&5\\
619614193*&1.890713176&13.59924378&102.56&13.0476&0.0028&30322&142&5.47 &0.03 &-1.74 &0.02 &sdB4.5&He9&VII&1.45 &9&7\\
593009050&2.102156123&49.08386838&17.46&15.8888&0.0028&29448&307&5.45 &0.05 &-2.70 &0.07 &sdB3.5&He4&VII&1.35 &4&7\\
66613068&2.189099272&12.2887394&15.19&14.6017&0.0028&33112&136&5.94 &0.03 &-2.17 &0.02 &sdO5&He6&VII&0.50 &6&7\\
364510227&2.630744589&20.00762956&137.2&11.5945&0.0028&35835&126&5.17 &0.04 &-1.20 &0.04 &sdB3&He11&IV&1.30 &11&4\\
689110148&2.718482764&26.50019871&13.41&16.9568&0.0029&28834&435&5.51 &0.05 &-2.43 &0.05 &sdB4&He6&VII&1.40 &6&7\\
364410161&2.923514276&19.34041113&48.68&14.7261&0.0028&28950&196&5.53 &0.03 &-2.80 &0.03 &sdB3.5&He4&VII&1.35 &4&7\\
593007055&2.936430799&46.80188977&56.06&14.3522&0.0028&45717&206&5.56 &0.03 &1.44 &0.12 &sdO9&He40&III&0.90 &40&3\\
248804155&3.670236008&8.064400154&16.35&16.4206&0.0028&-&-&-&-&-&-&sdO8&He12&V&0.80 &12&5\\
679405182&4.129512947&31.96125217&60.68&15.373&0.0029&28118&325&5.22 &0.04 &-1.89 &0.03 &sdB4&He8&VII&1.40 &8&7\\
615603055&4.216502336&52.14651705&10.97&16.7804&0.0028&31383&234&5.70 &0.06 &-2.51 &0.11 &sdB1&He6&VII&1.10 &6&7\\
615605186&4.230547616&51.2304863&14.09&16.3279&0.0038&33816&268&5.63 &0.05 &-2.92 &0.04 &sdO4&He6&VI&0.40 &6&6\\
593013047&4.625191224&48.80543938&11.41&15.2811&0.0028&26708&519&5.07 &0.05 &-1.52 &0.02 &sdB3&He11&VI&1.30 &11&6\\
266708166*&4.681327299&1.023185191&44.3&14.8603&0.0028&28264&500&5.38 &0.05 &-2.66 &0.05 &sdB0&He2&VII&1.00 &2&7\\
615603207&4.7865353&52.51187588&10.1&17.1244&0.0028&52767&1480&5.28 &0.08 &-1.40 &0.08 &sdO4&He11&V&0.40 &11&5\\
475310178&4.822308284&40.5022868&14.67&16.3149&0.0028&29105&410&5.33 &0.07 &-2.80 &0.09 &sdB2&He4&VII&1.20 &4&7\\
282604062&4.835819603&46.55188433&68.78&14.2451&0.0028&25049&0&5.42 &0.03 &-2.92 &0.07 &sdB6&He0&VII&1.60 &0&7\\
90110033&5.353294541&40.48253124&18.69&15.495&0.0028&28064&375&5.46 &0.06 &-2.55 &0.06 &sdB4&He5&VII&1.40 &5&7\\
163101192&5.764580014&24.3189442&27.79&15.4435&0.0028&38320&287&5.60 &0.04 &-1.54 &0.04 &sdB1&He10&VI&1.10 &10&6\\
615605166&5.863290213&51.13051918&20.88&15.9307&0.0028&26874&524&5.35 &0.05 &-2.63 &0.06 &sdB3&He6&VII&1.30 &6&7\\
90103101&5.980131673&42.15152611&18.43&15.7718&0.0028&29032&477&5.12 &0.06 &-2.76 &0.07 &sdB0&He6&VII&1.00 &6&7\\
605908160&6.035242726&56.02753697&21.25&16.1553&0.0028&36247&198&5.98 &0.03 &-1.54 &0.03 &sdO9.7&He20&VI&0.97 &20&6\\
689109217&6.137592106&26.81947135&30.48&16.876&0.0029&48808&323&5.85 &0.03 &0.77 &0.03 &sdOC8&He39&V&0.80 &39&5\\
248813198&6.365655979&8.965040771&15.91&15.5843&0.0028&30221&254&5.39 &0.05 &-2.79 &0.05 &sdO9&He3&VII&0.90 &3&7\\
284601234&6.43960764&31.59150803&37.67&15.5067&0.0028&27087&322&5.39 &0.03 &-2.20 &0.03 &sdB2.5&He7&VII&1.25 &7&7\\
360303217&6.473592337&30.08671271&51.55&15.0575&0.0029&35215&243&5.33 &0.04 &-0.87 &0.03 &sdB0&He14&V&1.00 &14&5\\
679407014&6.510622889&31.10567983&83.02&14.8487&0.0028&31036&91&5.67 &0.02 &-4.20 &0.34 &sdB0&He1&VII&1.00 &1&7\\
382607056&6.52831474&44.55505993&13.12&17.2211&0.0029&37229&281&5.97 &0.05 &-1.66 &0.06 &sdO9.7&He12&VII&0.97 &12&7\\
468710180&6.949149772&34.6740183&51.97&15.7564&0.0028&34935&121&5.85 &0.02 &-0.90 &0.01 &sdO9.5&He18&VII&0.95 &18&7\\
182909073&7.287589796&4.939829003&19.9&14.961&0.0029&35355&173&5.70 &0.02 &-3.65 &0.23 &sdO5&He2&VI&0.50 &2&6\\
615609066&7.329280389&52.97545911&11.52&16.7666&0.0028&42205&773&5.82 &0.07 &-4.15 &0.29 &sdO3&He2&V&0.30 &2&5\\
769504005&7.94839507&38.56917229&12.25&17.3717&0.0029&32778&377&6.30 &0.10 &-1.50 &0.08 &sdB2&He17&VII&1.20 &17&7\\
157510209&8.944011954&26.91505417&35.52&14.2571&0.0028&28583&226&5.65 &0.03 &-2.61 &0.04 &sdB3.5&He6&VII&1.35 &6&7\\
769508110&9.166012018&37.93184878&41.35&14.1229&0.0028&38970&1330&5.89 &0.10 &-3.20 &0.00 &sdO3&He2&VIII&0.30 &2&8\\
468705102&9.507188025&34.53228045&126.36&13.8805&0.0028&42088&205&5.43 &0.04 &-0.04 &0.02 &sdON9&He31&V&0.90 &31&5\\
472206190&9.522662426&43.74733706&12.44&16.1144&0.0029&37953&275&6.13 &0.05 &-1.53 &0.04 &sdB0.2&He11&VII&1.02 &11&7\\
250216131&10.52556586&5.15642968&62.46&12.784&0.0028&-&-&-&-&-&-&sdB1&He2&VI&1.10 &2&6\\
285308094&11.70129352&19.93682444&39.36&14.9629&0.0028&32891&111&5.84 &0.02 &-2.24 &0.03 &sdB1.5&He8&VII&1.15 &8&7\\
186304225&11.91879166&31.39871805&45.99&14.8024&0.0028&31069&166&5.65 &0.03 &-2.31 &0.04 &sdB2.5&He7&VII&1.25 &7&7\\
842609214&11.9977733&3.62908371&112.38&12.346&0.0028&37175&829&5.27 &0.09 &-3.27 &0.26 &sdO8&He1&VI&0.80 &1&6\\
164903013&12.32147759&20.9445829&27.56&14.5431&0.0028&27520&500&5.55 &0.07 &-2.48 &0.23 &sdB2&He8&VII&1.20 &8&7\\
14604029&12.45526919&35.36693272&29.95&14.8104&0.0029&36551&243&5.74 &0.04 &-1.48 &0.03 &sdB0&He14&VI&1.00 &14&6\\
186604095&12.73186541&43.56167204&65.84&14.5822&0.0029&-&-&-&-&-&-&sdB7&He7&VI&1.70 &7&6\\
870210221&12.77932243&0.708993387&20.6&15.8456&0.0028&38500&300&5.83 &0.07 &-1.00 &0.10 &sdO8&He15&VII&0.80 &15&7\\[0.5ex] 
\hline 
\end{tabular}
}
\label{tab:2}
\end{sidewaystable}
\twocolumn

\begin{figure*}[ht]
\centering
    \includegraphics[width=50mm]{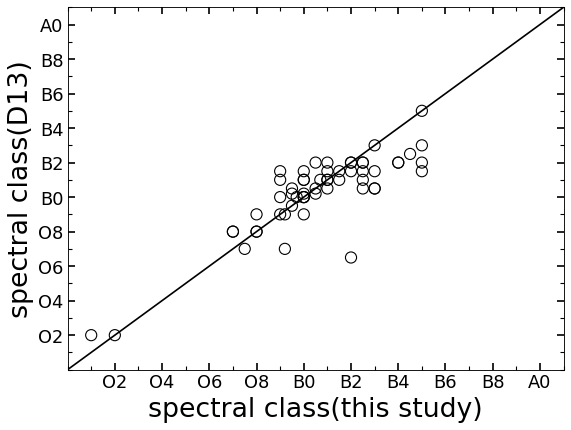}
    \includegraphics[width=50mm]{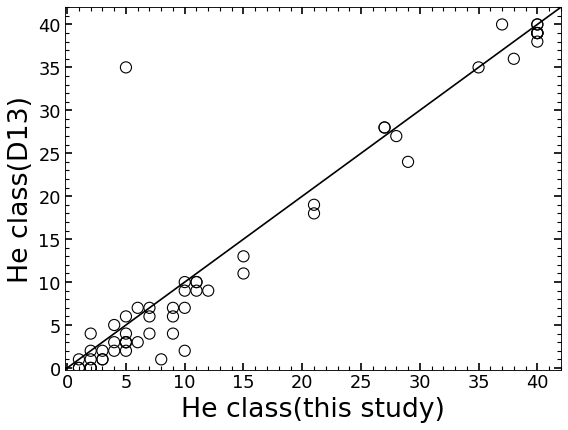}
    \includegraphics[width=50mm]{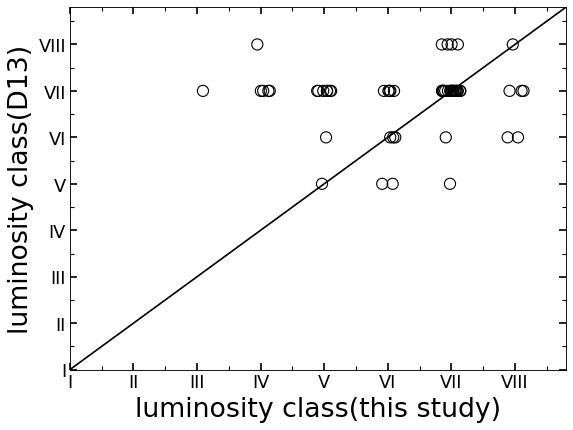}\\
    \includegraphics[width=50mm]{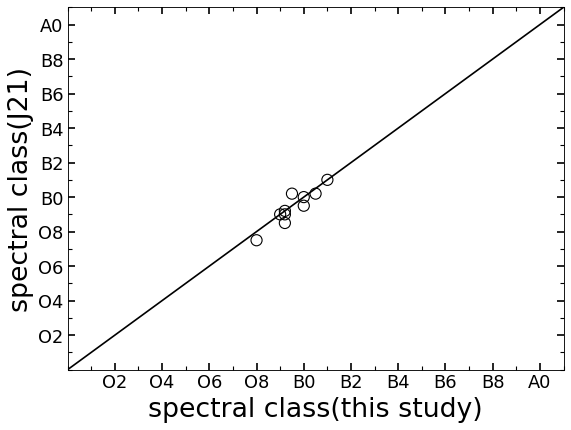}
    \includegraphics[width=50mm]{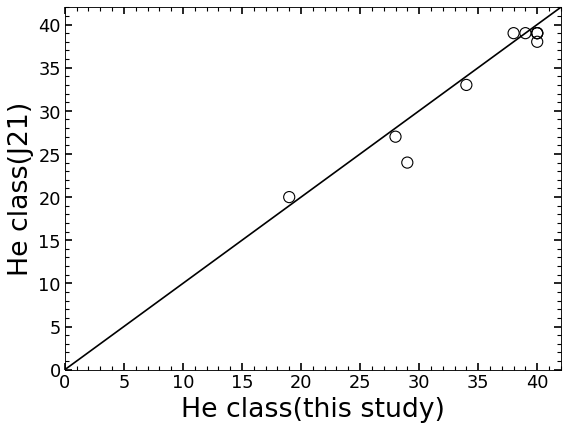}
    \includegraphics[width=50mm]{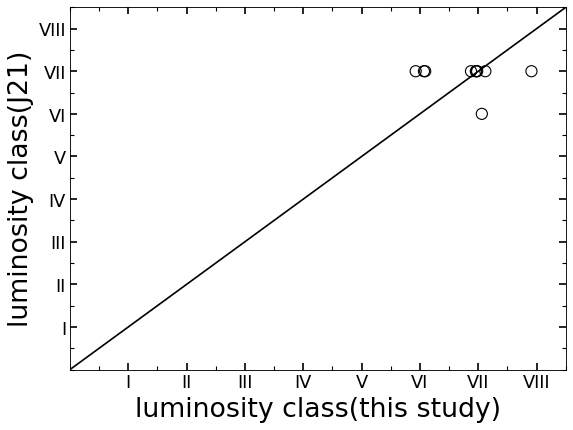}    
    \caption{Spectral classification comparison with other studies. Top panels  for comparison with D13, while bottom panels for comparison with \citet{2021MNRAS.501..623J}.}
    \label{fig6}
\end{figure*}

\subsection{Comparison with other studies}

In order to define the MK-like spectral classification system, D13 analyzed 181 hot subdwarfs and blue horizontal branch (BHB) stars from the sample of \citet{1986ApJS...61..305G}, among which 76 of them were selected as standard stars. Furthermore, using MK-like classification system, \citet{2021MNRAS.501..623J} classified 107 He-rich hot subdwarf stars observed by SALT.  We cross-matched the hot subdwarfs analyzed in this study with the sample analyzed in D13 and \citet{2021MNRAS.501..623J}, and found 57 and 10 common objects, respectively. 

Fig \ref{fig6} shows comparison of spectral classification between this study and the two studies mentioned above (e.g., top panels for comparison with D13, and bottom panels for comparison with \cite{2021MNRAS.501..623J}). It is clearly presented a good consistent for both spectral class and He class in these studies. There is an  exception object (i.e., LAMOST 880808240 = PG 1300+279) in He class comparison with D13. It was assigned a number of 5 for He class in this study that denotes an H-rich atmosphere. While it was assigned a number of 35 in D13 that denotes an He-rich atmosphere. We checked this object carefully, and found dominated H lines in the LAMOST spectrum. The reason for this conflict could be the different spectral quality from different observation instruments. The systematic errors of spectral class and He class between our results and D13 are 0.098 and 1.4 respectively, while the systematic errors of spectral class and He class between our results and \citet{2021MNRAS.501..623J}  are 0.029 and 2.37.

On the other hand, the comparison for luminosity class present much wider dispersion between these studies. The systematic errors for luminosity class between our results and D13 is 0.98, and is 0.5 when comparing with  \citet{2021MNRAS.501..623J}. 
Since luminosity class is obtained by comparing the line width with the average line width of all sample (see Section 2.2), thus it could be somewhat subjective and result in wider dispersion between different studies. Actually, \citet{2021MNRAS.501..623J} also obtained a wide dispersion of luminosity class when comparing their results with that of D13 (see Fig B.1 in their study). 

\subsection{The relationships between spectral classification and atmospheric parameters.}

\begin{figure*}
\centering
    \includegraphics[width=56mm]{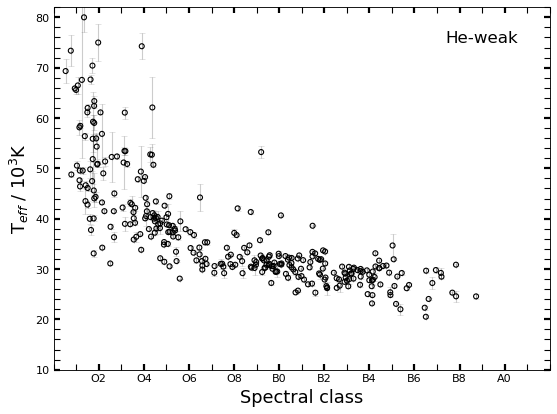}
    \includegraphics[width=56mm]{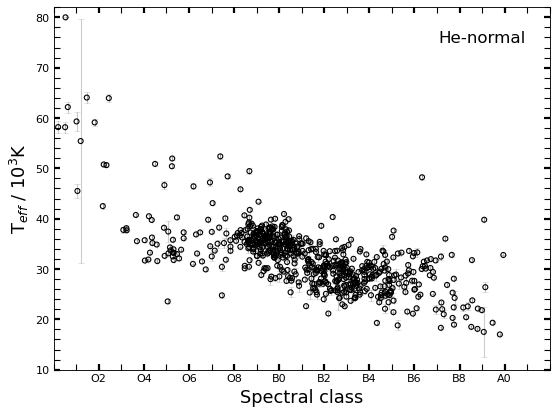}
    \includegraphics[width=56mm]{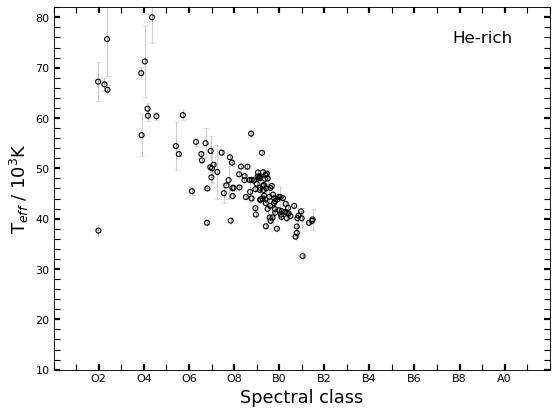}
   \caption{Spectral class - $T_{\rm eff}$ plane. A uniform jitter of $\pm$ 0.05 was added into the spectral class for clarity.}
   \label{fig7}
\end{figure*}

Since spectral classification is mainly based on stellar spectra and independent of atmospheric models, thus it is possible to verify atmospheric parameters by spectral classification (D13, \cite{2021MNRAS.501..623J}).
Among 1224 hot subdwarfs classified in this study, 1084 of them have atmospheric parameters (e.g., $T_{\rm eff}$, $\log{g}$ and $\log(n{\rm He}/n{\rm H})$) reported in \citet{2022A&A...662A..40C}. We presented the relationships in this section between spectral classification and atmospheric parameters for these stars. 

\begin{figure}
	\includegraphics[width=80mm]{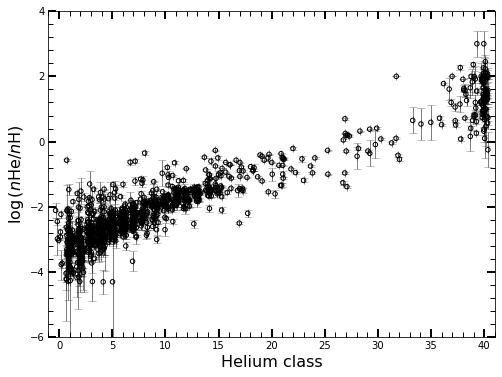}
    \caption{He class - $\log(n{\rm He}/n{\rm H})$ plane. A uniform jitter of $\pm$ 0.15 was applied to Helium class for clarity.}
    \label{fig8}
\end{figure}

The panels from left to right in Figure \ref{fig7} shows the relationship between spectral class and effective temperatures (i.e., $T_{\rm eff}$) for He-weak, He-normal and He-rich (including He-strong and He-strong C) stars, respectively. As expected that, more earlier of the spectral class in each panel, much higher of the surface effective temperatures. This tendency is especially clear for stars with spectral class from A0 to O4. As for hot subdwarf stars with spectral class earlier than O4, $T_{\rm eff}$ show much wider dispersion with similar spectral class. This is due to the fact that these stars have very high temperatures (e.g., $\geq$ 40000 K), thus have similar spectral properties  which lead to similar classification of spectral class. Moreover, the uncertainties also become larger with higher $T_{\rm eff}$ values (see gray error bars in the figure).  

One also can see from Fig 7 that, all the He-rich stars (e.g., right panel) present spectral class earlier than B2, while He-weak (left panel) and He-normal (middle panel) stars have wide spectral class distribution from A0 to O1. This is due to the fact that, HeI and /or HeII lines dominated spectra basically require higher effective temperatures than H Balmer lines dominated spectra, thus present more earlier spectral class. On the whole, these results demonstrate appropriate spectral class assigned for the selected samples. 

\begin{figure*}
\centering
    \includegraphics[width=56mm]{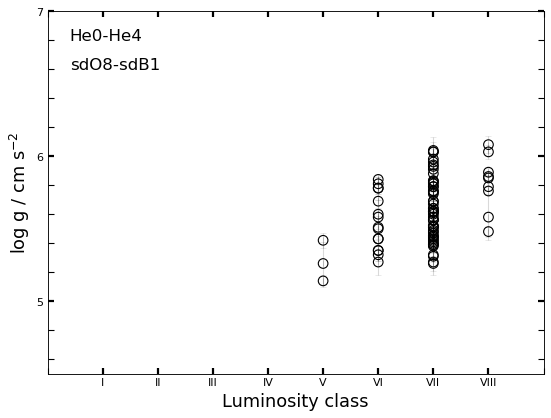}
    \includegraphics[width=56mm]{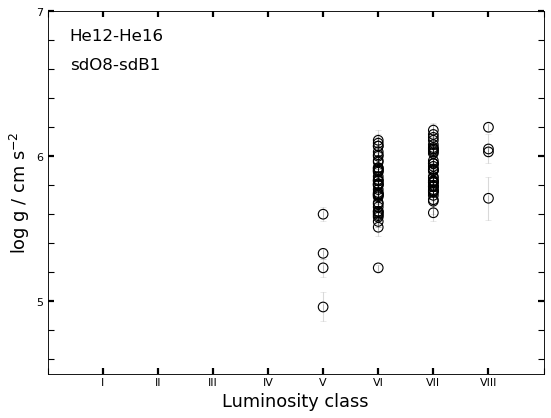}
    \includegraphics[width=56mm]{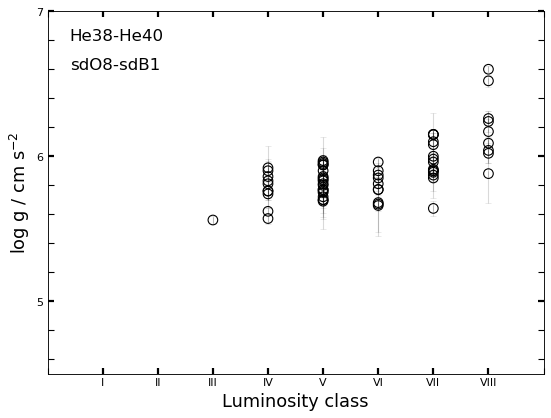}
    \caption{Luminosity class - $\log{g}$ plane. To show the relationship between luminosity class and $\log{g}$ clearly, hot subdwarf stars in all panels are selected within relative narrow range of spectral class and He class. See context for detailed description.}
    \label{fig9}
\end{figure*}

Fig \ref{fig8} presents the relationship between He class and surface He abundance (i.e., $\log(n{\rm He}/n{\rm H})$). As one can see in the figure, with the integer number of He class increase from 0 to 40, the surface He abundance also increase gradually from $\log(n{\rm He}/n{\rm H}) \approx$ -4 (e.g., pure H envelope) to $\log(n{\rm He}/n{\rm H}) \approx$ 3.5 (e.g., pure He envelope), which indicates a good consistency for the two independent parameters.

Since atmospheric parameters (e.g., effective temperature, surface gravity and He abundance) would influence line widths of the spectra, thus influence the classification of luminosity class. Therefore, to show the relationships between luminosity class and surface gravity more clearly, the 3 panels in Fig \ref{fig9} present luminosity - $\rm{log}\ g$ plane for hot subdwarf stars within relative narrow ranges of  spectral class (i.e., sdO8 - sdB1) and He class (i.e., He0 - He4 for left panel, He12 - He16 for middle panel, and He38 - He40 for right panel).

Most of hot subdwarf stars in Fig \ref{fig9} have luminosity class from V to VIII, while a few of them have luminosity class from II to IV. As expected,  with $\rm{log}\ g$ increasing in all the panels, luminosity classes of the selected hot subdwarfs also increase, which denotes that luminosity to mass ratios of the stars becomes lower. However, for each type of luminosity class (e.g., VII), hot subdwarf stars show a wide dispersion of $\rm{log}\ g$. It is due to the fact that, besides $\rm{log}\ g$, other parameters could affect line widths of the spectra as well, such as effective temperature, He abundance, rotation, etc. D13 also obtained a similar results (see Fig 14 in their study). Note that the sample analyzed in D13 are much less than the sample in this study. Most of stars in their study have luminosity class VII, while stars with other luminosity class are few. Further more, some BHB stars were included in D13,  thus more luminous luminosity classes  (e.g., I and II) also appeared in their sample.

\subsection{Fitting formula for spectral class}

Classification of spectral class based on visual inspection involves comparing depth and width of spectral  lines, thus it could be time-consuming, especially when dealing with a large number of spectra. We developed several formulae for spectral class by fitting linearly the classification results obtained in this study. These formulae can be used to classify new hot subdwarfs identified with LAMOST spectra automatically without visual inspection. 

\begin{figure*}
\centering
    \includegraphics[width=70mm]{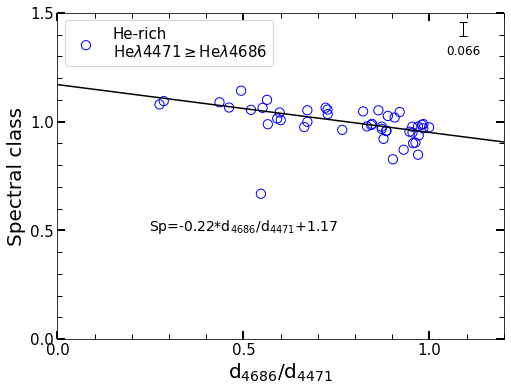}
    \includegraphics[width=70mm]{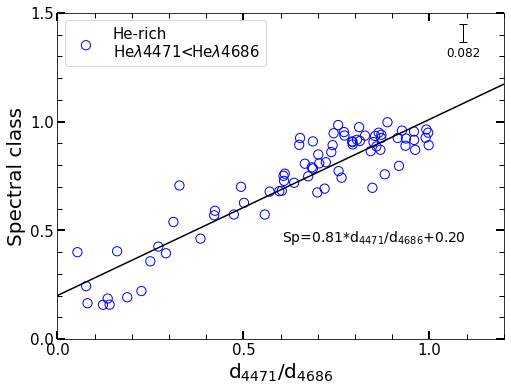}\\
    \includegraphics[width=70mm]{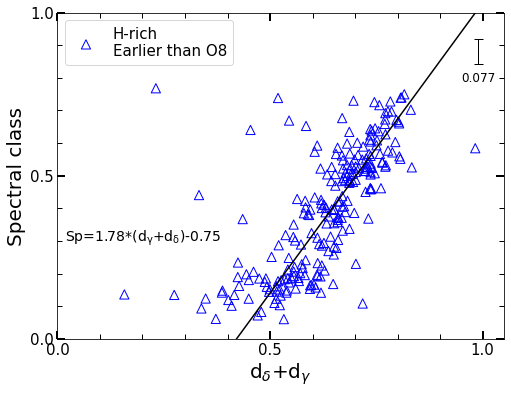}
    \includegraphics[width=70mm]{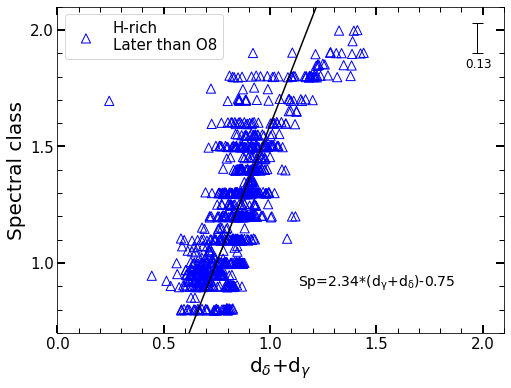}
    
    \caption{The relationship between spectral class and line depth. Top panels for He-rich hot subdwarfs, while bottom panels for H-rich hot subdwarfs (see context for more details). The black solid lines are the linear fitting lines of each population. A uniform jitter of ± 0.05 was added into the spectral class of each spectrum for clarity. The black vertical error bars in each panel are the root mean square differences between the fitting values and observed spectral classes. }
    \label{fig10}
\end{figure*}

Based on the strength of He lines, hot subdwarf spectra can be divided roughly into two primary categories: He-rich population (corresponding to He-strong and He-strong C category in D13, also see \cite{2021MNRAS.501..623J}) and H-rich population (corresponding to He-normal and He-weak category in D13). Spectral class for He-rich population can be fitted by two following equations. If line depth of HeI $\lambda\lambda4471$ is less than line depth of He II $\lambda\lambda 4686$: 
\begin{equation}
   Sp=0.81*d_{\rm 4471}/d_{\rm 4686}+0.20, 
	\label{eq:5}
\end{equation}
where Sp corresponds to a numerical scale for spectral class, e.g., spectral class O2 = 0.2, O5 = 0.5, B0 = 1.0, etc. While $d_{\rm 4471}$ and $d_{\rm 4686}$ denote line depth of HeI $\lambda\lambda 4471$ and HeII  $\lambda\lambda 4686$, respectively. 

If line depth of HeI $\lambda\lambda 4471$ is greater than depth of HeII $\lambda\lambda 4686$: 
\begin{equation}
    Sp=-0.22*d_{\rm 4686}/d_{\rm 4471}+1.17.
	\label{eq:6}
\end{equation}

For H-rich population, there are also two fitting formulae which can be used to obtain the spectral class. If the spectral class earlier than O8:
\begin{equation}
Sp=1.78*(d_{\rm \gamma}+d_{\rm \delta})-0.75,
\label{eq:7}
\end{equation}
where $d_{\gamma}$ and $d_{\rm \delta}$ are line depths for $\rm{H}_{\gamma}$ and $\rm{H}_{\delta}$, respectively. If the spectral class later than O8:
\begin{equation}
Sp=2.34*(d_{\rm \gamma}+d_{\rm \delta})-0.75.
\label{eq:8}
\end{equation}

The relationships between spectral class and the depths of spectral lines are shown in Fig \ref{fig10}. The fitting lines (black solid line) based on the formulae described above for each population are also shown. It is important to note that though fitting formulae can improve the spectral classification efficiency, it may not capture all details of spectral lines. Therefore, visual examination can still be valuable to validate and refine the results obtained through quantitative analysis.

\section{Conclusions}
Employing the MK-like spectral classification scheme designed by D13, we classified 1224 hot subdwarf stars with LAMOST DR9 low-resolution spectra. To obtain specific spectral class, luminosity class and He class for all sample, standard stars were selected from LAMOST. Among the 1224 classified hot subdwarf stars, 702 stars belong to He-normal category, 396 stars  belong to He-weak category, 81 stars  belong to He-strong C category, and 45 stars belong to He-strong category.

A good consistency between spectral classification and atmospheric parameters (e.g., effective temperature, gravity and He abundance) indicates appropriate spectral classification in this study. Linear fitting formulae for spectral class were obtained by fitting the distribution of data in line depth vs spectral class plane.

\section*{Acknowledgements} 
We thank  the referee for valuable suggestions and comments which helped improve the manuscript greatly. This work acknowledges support from the National Natural Science Foundation of China (No. 12073020 ), Scientific Research Fund of Hunan Provincial Education Department grant No. 20K124, Cultivation Project for LAMOST Scientific Payoff and Research Achievement of CAMS-CAS, the science research grants from the China Manned Space Project with No. CMS-CSST-2021-B05. Guoshoujing Telescope (the Large Sky Area Multi-Object Fiber Spectroscopic Telescope LAMOST) is a National Major Scientific Project built by the Chinese Academy of Sciences. Funding for the project has been provided by the National Development and Reform Commission. LAMOST is operated and managed by the National Astronomical Observatories, Chinese Academy of Sciences.

\end{document}